\documentclass[conference,10pt]{IEEEtran}

\newcommand{\MATLAB}{\textsc{Matlab}\xspace}

\usepackage[utf8]{inputenc}
\usepackage[T1]{fontenc}
\usepackage{textcomp}
\usepackage{fixltx2e}
\usepackage{multirow}

\usepackage[pdftex]{graphicx}
\DeclareGraphicsExtensions{.pdf,.png,.jpg}
\usepackage[caption=false,font=footnotesize]{subfig}
\usepackage{booktabs}
\usepackage{url}
\usepackage{amsmath}
\usepackage{amssymb}
\usepackage{amsthm}
\usepackage{acronym}
\usepackage[capitalise,noabbrev]{cleveref}
\usepackage[textwidth=1cm]{todonotes}
\usepackage{xspace}
\usepackage{booktabs}
\usepackage[textwidth=1cm]{todonotes}
\usepackage[american]{babel}
\usepackage[color]{changebar}
\usepackage{textgreek}
\usepackage{enumitem}
\usepackage[bottom]{footmisc}
\usepackage{mathrsfs}

\usepackage{csquotes}
\usepackage[backend=biber,style=ieee,doi=false,isbn=false,mincitenames=1,maxcitenames=2]{biblatex}
\DeclareFieldFormat{sentencecase}{#1} 
\DeclareFieldFormat{titlecase}{#1} 
\usepackage{xpatch}
\xpatchbibmacro{textcite}{\addspace}{\addnbspace}{}{}
\xpatchbibmacro{Textcite}{\addspace}{\addnbspace}{}{}

\DefineBibliographyStrings{english}{
	andothers = et~al\adddot\addspace
}

\def\todoCtd#1{%
TODO: #1%
\ifx&#1&...\fi%
\endgroup
\cbend
\relax
}

\usepackage[load-configurations=binary,detect-all,binary-units=true,range-phrase=--,per-mode=symbol,range-units=single,list-units=single]{siunitx}

\DeclareSIUnit\dBm{dBm}
\DeclareSIUnit\dB{dB}

\acrodef{OFDM}{orthogonal frequency division multiplexing}
\acrodef{MCS}{modulation and coding scheme}
\acrodef{BPSK}{binary phase shift keying}
\acrodef{CP}{cyclic prefix}
\acrodef{LS}{least square}
\acrodef{CIR}{channel impulse response}
\acrodef{CFR}{channel frequency response}
\acrodef{CSI}{channel state information}
\acrodef{FFT}{fast Fourier transformation}
\acrodef{IFFT}{inverse FFT}
\acrodef{LTF}{Long Training Field}
\acrodef{VHT-LTF}{very high throughput LTF}
\acrodef{mmWave}{millimiter-wave}
\acrodef{V2V}{vehicle-to-vehicle}
\acrodef{OMP}{orthogonal maching pursuit}
\acrodef{PDD}{packet detection delay}
\acrodef{SFO}{sampling frequency offset}
\acrodef{CFO}{carrier frequency offset}
\acrodef{AWGN}{additive white Gaussian noise}
\acrodef{AND}{atomic norm denoising}
\acrodef{MPC}{multipath component}
\acrodef{SNR}{signal to noise ratio}
\acrodef{L-LTF}{legacy long training field}
\acrodef{VHT-LTF}{very high throughput long training field}
\acrodef{LTF}{long training field}
\acrodef{3GPP}{3rd Generation Partnership Project}
\acrodef{NYUSIM}{NYUSIM}
\acrodef{QuaDRiGa}{QUAsi Deterministic RadIo channel GenerAtor}
\acrodef{BW}{channel bandwidth}
\acrodef{ToF}{time of flight}
\acrodef{MMSE}{minimum mean square error}
\acrodef{JCAS}{joint communication and sensing}
\acrodef{IoT}{internet of things}
\acrodef{OTFS}{orthogonal time frequency space}
\acrodef{PAPR}{peak-to-average-power ratio}
\acrodef{SDR}{software defined radio}

\begin{document}

\title{Practical Channel Splicing using OFDM Waveforms for Joint Communication and Sensing in the IoT}
\author{
\IEEEauthorblockN{Sigrid Dimce\IEEEauthorrefmark{1}, Anatolij Zubow\IEEEauthorrefmark{1}, Alireza Bayesteh\IEEEauthorrefmark{2},\\ Giuseppe Caire\IEEEauthorrefmark{3}, and Falko Dressler\IEEEauthorrefmark{1}}
\IEEEauthorblockA{\IEEEauthorrefmark{1} Telecommunication Networks, TU Berlin, Germany}
\IEEEauthorblockA{\IEEEauthorrefmark{2} Huawei, Canada}
\IEEEauthorblockA{\IEEEauthorrefmark{3} Communications and Information Theory, TU Berlin, Germany}
\texttt{\{dimce, zubow, dressler\}@tkn.tu-berlin.de, alireza.bayesteh@huawei.com,}\\
\texttt{caire@tu-berlin.de}
}

\maketitle

\begin{abstract}
Channel splicing is a rather new and very promising concept.
It allows to realize a wideband channel sounder by combining multiple narrow-band measurements.
Among others, channel splicing is a sparse sensing techniques suggested for use in \ac{JCAS}, channel measurements and prediction using cheap hardware that cannot measure wideband channels directly such as in the \ac{IoT}.
This work validates the practicality of a channel splicing technique by integrating it into an OFDM-based IEEE 802.11ac system, which we consider representative for many \ac{IoT} solutions.
Our system allows computing both the \ac{CIR} and the \ac{CFR}.
In this paper, we concentrate on the impact of the number of sub-bands in our study and show that even using only 50\% of the overall spectrum leads to very accurate \ac{CIR} measures.
We validate the system in simulation and confirm the results in an experimental in-door scenario using software defined radios.
\end{abstract}

\acresetall

%

\section{Introduction}
\label{sec:Intro}

\Ac{JCAS} is becoming more important in different application domains, both in 6G as well as for the \ac{IoT} \cite{wang2022integrated,zhang2022enabling,wild2021joint,cui2021integrating}.
This also triggered an ongoing discussion of appropriate waveforms, \ac{OTFS} being considered a possible compromise \cite{wu2021otfsbased}.
However, most existing communication systems are based on \ac{OFDM}, so, integrating sensing here is very important \cite{liyanaarachchi2020joint}.

Channel sounding is the core functionality required for \ac{JCAS}, but also for channel estimation and channel prediction techniques.
For example, commodity WiFi devices have been used for purposes other than wireless communication such as indoor localization and channel sounding due to the relatively low cost~\cite{vasisht2016decimeter, khalilsarai2020wifibased}.
Channel sounding relies on the \ac{CSI} of the communication link.
In case of a single-antenna device, the \ac{CSI} is equivalent to the \ac{CFR}, which provides channel information in the frequency domain. 
Through Fourier transformation, the channel is characterized in the delay domain, represented by the \ac{CIR}. 
The \ac{CIR} describes the multipath channel over the delay domain. 
However, the accuracy of the estimated \ac{CIR} is limited by the \ac{BW} supported by the system.
Precise measurement of multipath components is only possible using wideband signals. 
In the delay domain, the resolution is equal to $1/\mathit{BW}$, which, multiplied with the speed of light, gives the necessary difference between the distance of two distinct paths.
For instance, a bandwidth of \SI{20}{\mega\hertz} implies that the distance traveled by the signal from two distinct paths should have at least a difference of \SI{15}{\meter}, so that the paths are distinguishable at the receiver~\cite{khalilsarai2020wifibased}.

\begin{figure}
	\centering
    \includegraphics[width=\linewidth]{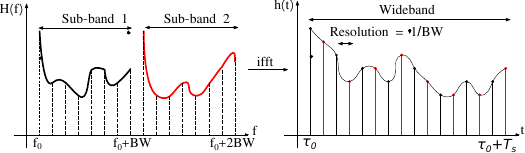} 
    \caption{Illustration of the channel splicing concept}
	\label{fig:intro}
\vspace{-.8em}   
\end{figure}

However, in general, wideband sounding is quite complex and energy inefficient for existing communication systems.
Also, many \ac{IoT} systems can only process narrow-band subchannels.
A possible solution to this issue is given by multi-band sensing or spectrum splicing \cite{wan2022fundamental}.
Conceptually, channel splicing means measuring multiple narrow-band sub-bands/subchannels and then combining the results to obtain, ideally, the same results as generated by a single wideband measurement.
The concept is depicted in \cref{fig:intro}.

Initially, spectrum splicing was developed targeting indoor localization applications \cite{khalilsarai2020wifibased,kazaz2021delay, xie2018precise}, and later extended to other applications such as human sensing~\cite{shen2019wirim}.
This technique allows a single WiFi device to transmit packets and extract \ac{CSI} in the multiple frequency bands tens of megahertz wide.
The collected information through multiple bands is combined mimicking a wideband channel. 
Channel splicing exploits the sparse nature of the \ac{CIR} and applies sparse recovery method on the collected data to obtain \ac{CIR} with high resolution. 
The delay associated with the shortest path enables estimating the signal \ac{ToF}, which is used for localization.
As we will demonstrate in this paper, splicing can also be performed on a subset of the sub-bands with only small reduction in accuracy.
Despite its advantages, splicing faces several challenges due to transceiver impairments. 

A few algorithms have been developed aiming to enhance the splicing performance and resolve the impairments challenges~\cite{khalilsarai2020wifibased, vasisht2016decimeter, kazaz2021delay,wan2022fundamental,xie2018precise, wan2022multiband}. 
Our work builds upon the theoretical concepts of spectrum splicing presented in~\cite{khalilsarai2020wifibased}. 
We first implement a communication system in \MATLAB{} based on the IEEE 802.11ac standard. 
We perform channel estimation based on the \ac{LS} estimation technique both in time and frequency domain. 
We extended the system by spectrum splicing allowing to combine \ac{CFR} measurements from multiple bands. 
The channel sounder utilizes \ac{SDR} components, we used two USRP N310 in the lab, for transmitting/receiving the signal over the air.
We validate the practical use of splicing technique in simulations and indoor experiments by splitting a wide channel bandwidth into narrower bands, collect the \ac{CFR} over these bands, perform splicing on the collected data and compare the estimated \ac{CIR} with the wide channel.
Our results build the basis for \ac{OFDM}-based \ac{JCAS} solutions and for low-cost \ac{IoT} channel sounding.

Our main contributions can be summarized as follows:
\begin{itemize}
    \item We implement a practical \ac{OFDM} communication system based on IEEE 802.11ac and the \ac{LS} estimation technique on the receiver to compute both \ac{CIR} and \ac{CFR}.
    \item We extend the system by incorporating the spectrum splicing to investigate the impact of the number of sub-bands, and the accuracy of the sensing approach.
    \item We validate the splicing technique in a controlled simulation environment as well as in a set of indoor experiments focusing on the numbers of \acl{MPC}.
\end{itemize}

%

\section{Related Work}
\label{sec:related}

Channel sounding is a crucial technique for generating knowledge to characterize the wireless channel in a certain frequency band.
The principle of sounding is to transmit a known baseband signal up-converted at the frequency of interest, which is post-processed on the receiver side for extracting metrics (\ac{CIR}, \ac{CFR}) that provide channel information.
Several sounding techniques were developed over the years with the sole purpose to accurately characterize the propagation channel.
The two most common techniques are the spread spectrum sliding correlator~\cite{rappaport1996wireless} and the \ac{OFDM}-based system~\cite{heiskala2001ofdm}. 
Both techniques have demonstrated undeniable success at lower frequencies, as well as at \ac{mmWave} frequency band~\cite{sun2017design, lv2019channel}.
The OFDM-based system consists of multiple subcarriers, where each subcarrier sees the channel as flat, hence, allowing it to overcome the main drawback of wideband transmission in terms of frequency selectivity.
Nevertheless, the system requires a large \ac{PAPR} as well as tight receiver synchronization. 
On the other hand, the spread spectrum sliding correlator ``spreads'' the carrier signal over a large bandwidth by mixing it with a binary PN sequence. 
The received signal is mixed with a slower identical version of the PN sequence, which makes the system less vulnerable towards interference, but more complex in terms of hardware and software implementation compared to the \ac{OFDM} system. 
Thus, choosing one of the aforementioned sounding techniques is application specific and requires to compromise between robustness and complexity.

Propagation information is provided in the delay and frequency domain, respectively, \ac{CIR} and \ac{CFR}, which are computed by the channel estimation techniques.
Among the existing ones, the \ac{LS} estimation is the most common method characterized by low computational complexity.
Yet, in a few application scenarios, this technique yields inferior performance~\cite{le2021deep}.
Another well-known estimation method is the \ac{MMSE}~\cite{heiskala2001ofdm}, which minimizes the channel estimation error.
However, \ac{MMSE} leads to high computational complexity and requires prior channel statistic information, which sometimes is not available.
Beside these traditional techniques, new models based on deep learning are being developed~\cite{le2021deep}, targeting performance improvement, and using \ac{LS} estimation for training their model. 

Reducing system costs by using existing hardware infrastructure for applications such as channel sounding is becoming more crucial.
In this aspect, spectrum splicing methods are attracting high attention in the recent years.
Most of the existing works target indoor localization applications~\cite{vasisht2016decimeter, wan2022multiband, kazaz2021delay, khalilsarai2020wifibased}, and a few others e.g., human sensing~\cite{shen2019wirim}. 
The developed methods aim estimating the \ac{MPC} delays precisely considering the present of hardware distortions. 
For instance, Chronos~\cite{vasisht2016decimeter} is an indoor positioning system, which estimates sub-nanosecond \ac{ToF} using compressed sensing sparse recovery methods on the collected \acp{CFR} over multiple bands. 
The algorithm addresses the phase offset issue as a result of hopping between frequency bands. 
Splicer~\cite{xie2018precise} is a software-based system that splices  multiple \acp{CSI} measurements and achieves single-AP localization. 
The authors propose several techniques for hardware impairment corrections.
The work in~\cite{wan2022multiband} presents a two-stage global estimation scheme, which aims to improve estimation accuracy firstly by achieving initial delay estimation based on a coarse signal model and then a global delay estimation.
Other authors~\cite{kazaz2021delay} use the shift-invariance structure in the multiband \ac{CSI} and propose a weighted gridless subspace fitting algorithm for the delay estimation. 
The fundamental limits and optimization of multi-band splicing in terms of the time delays are analyzed and presented in~\cite{wan2022fundamental}.
The statistical resolution limit is derived for the delay resolution and an algorithm is proposed to solve the parameters optimization problem.
A grid-based multi-band splicing technique is presented in~\cite{khalilsarai2020wifibased}.
The method is characterized by low-complexity and can easily scale to large-dimensional problems.
Besides localization, other mechanisms, such as WiRIM~\cite{shen2019wirim}, utilize spectrum splicing for human sensing applications.

We integrate the theoretical concepts proposed in~\cite{khalilsarai2020wifibased} into an \ac{OFDM}-based \ac{JCAS} system.
We validated our practical implementation of channel splicing both in simulation as well as an experimental indoor scenario.

%

\section{Spectrum Splicing Architecture}
\label{sec:splicing}

Our channel sounder is based on the spectrum splicing technique presented in~\cite{khalilsarai2020wifibased}. 
The developed method exploits the sparse nature of the \ac{CIR} and utilizes a grid-based compressed sensing technique to estimate the path delays and amplitude.
On a high abstraction level, the idea is to use subchannels to reconstruct the \ac{CIR} with high resolution as shown in \cref{fig:intro}.

\subsection{Spectrum Splicing Concept}
\label{sec:ss_concept}

The system is based on \ac{OFDM} and packets are transmitted over \textit{M} frequency bands, where each band is composed of \textit{N} subcarriers. 
The subcarriers are indexed according to the integer set $\mathcal{N} = \{- \frac{N-1}{2}, ..., \frac{N-1}{2} \}$,
where $N$ is an odd integer.
The presented technique performs the estimation based on the pilot signals, which in the receiver side can be written as
\begin{equation}
	y[m,n] = H[m,n]S_{m,n} + z[m,n], \quad m \in [M], n \in \mathcal{N}
\end{equation}
where $S_{m,n}=1$ is the pilot symbol transmitted over the \textit{n}-th subcarrier of band $m$ (\textit{$f_{m,n}$}), which without loss of generality is assumed to be 1; $H[m,n]$ denotes the \ac{CFR} in the same subcarrier, and the $z[m,n] \sim \mathcal{CN}(0,\,1/\acs{SNR})$ is the \ac{AWGN} channel.

Assuming that the propagation environment is comprised of  \textit{K} scatters, the \ac{CIR} over the delay domain is given as
\begin{equation}
	h(\tau) = \sum_{k=1}^{K}c_{k}\delta(\tau - \tau_{k}),
\end{equation}
where $\delta (\cdot)$ stands for Dirac's delta function, $\tau_{k} \in [0,1/f_{s})$ is the delay associated with each path $k$ ($f_{s}$ - subcarrier spacing), and $c_{k} \in \mathbb{C}$ is corresponding amplitude. 
The parameter of gain and delay are independent of the frequency band. 
On the other hand, the \ac{CFR} samples are computed from the \ac{CIR} via the Fourier transformation, and can be written as
\begin{equation}
	H[m,n] = \mathscr{F}\{h(\tau)\}\mid_{f=f_{m,n}} = \sum_{k=1}^{K}c_{k}e^{-j2\pi f_{m,n}\tau_{k}},
\end{equation} 
for $m \in [M], n \in \mathcal{N} $.
Furthermore, during the transmission, the pilot samples are affected by several distortions caused by the hardware devices, which lead to the phase term
\begin{equation}
	\psi[m,n] = -2\pi(\delta_{m}nf_{s} + \phi_{m}), \quad m \in [M], n \in \mathcal{N}
\end{equation}
where $\delta_m \in [0,1/f_{s})$ stands for the timing offset due to the \ac{PDD} and the \ac{SFO}, and $\phi_{m} \in [0,1)$ represents the phase offset due to the \ac{CFO} between transmitter and receiver, and the phase offset due to switching channel bands. 
The two parameters, $\delta_m, \phi_{m}$ differ from one band to the other such that, $\psi[m,n]$ is a linear function in each band, of the subcarrier index with different slope and constant term.
As a result, the received pilot samples, including the distortion component are represented in the following form
\begin{equation}\label{eq:received_samples}
    \begin{aligned}
		y[m,n] & = e^{j\psi[m,n]}H[m,n] + z[m,n]\\ 
		& = e^{j\psi[m,n]}\sum_{k=1}^{K}c_{k}e^{-j2\pi f_{m,n}\tau_{k}} + z[m,n]\\
		& = \sum_{k=1}^{K}c_{k}e^{-j2\pi(f_{m,0}\tau_{k} + \phi_{m})}e^{-j2\pi nf_{s}(\delta_{m} + \tau_{k})} + z[m,n]		 
    \end{aligned}
\end{equation}
where for each band the subcarriers are assumed to be equispaced with a space equal to $f_{s}$  and $f_{m,n} = f_{m,0} + nf_{s}, n \in \mathcal{N}$, and $f_{m,0}$ being the carrier frequency of band $m$.

The proposed spectrum splicing technique aims to estimate the \ac{CIR} based on noisy and distorted pilot samples, using the following steps:
\begin{enumerate}
	\item For each band $m \in [M]$ estimate and remove the distortion parameters {$\delta_{m}, \phi_{m}$}.
	The estimation is performed using the sparse recovery technique of \ac{AND}.
	\item Splice the clean pilot data to obtain a high-resolution estimated \ac{CIR} using the \ac{OMP} sparse recovery technique.
	\item Resolve ambiguities using a hand-shaking procedure between the two communication nodes.
\end{enumerate}

In this work, we focus on the second step of multi-band splicing and perform the estimation based on the \ac{VHT-LTF} of 802.11ac frame. 
On the receiver side, time and \ac{CFO} is estimated and corrected. 


The multi-band splicing technique consists of merging the measurements conducted over several bands, increasing the resolution of the estimated \ac{CIR} by expanding the measurement bandwidth. 
For instance, the resolution over the delay domain obtained from the measurements over a single band is given as $(\Delta \tau)_{1} = 1/Nf_{s}$, whereas over $M$ frequencies band it increases to $(\Delta \tau)_{1} = 1/MNf_{s}$. 
Considering that the \ac{CIR} is sparse, the authors in~\cite{khalilsarai2020wifibased} used compressed sensing, and more specifically the \ac{OMP} sparse recovery method to recover the \ac{CIR}. 
Firstly, a vector is defined containing the subcarriers per band \textbf{f}($m$) = $[f_{m,-(N-1)/2}, ...,f_{m,(N-1)/2}]^{T}$
and for all the bands $\textbf{f} =[\textbf{f}(1)^{T},..., \textbf{f}(M)^{T}]^{T} \in \mathbb{R}^{MN} $.
Similarly, a vector is defined containing the clean \ac{CFR} samples per band $\tilde{\textbf{y}}(m) = [\tilde{y}[m,-(N-1)/2],..., \tilde{y}[m,(N-1)/2]]^{T} $ and for all the bands $\tilde{\textbf{y}} = [\tilde{\textbf{y}}(1)^{T},..., \tilde{\textbf{y}}(M)^{T}]^{T} \in \mathbb{C}^{MN}$.
Furthermore, the elements of the $\tilde{\textbf{y}}$ can be written as
\begin{equation}\label{eq:ss_received_pilots}
	[\tilde{\textbf{y}}]_{i} = \mathscr{F}\{h_{0}(\tau)\}\mid_{[\textbf{f}]_{i}} + [ \tilde{\textbf{z}}_{i}]\, \quad i = 1,...,MN 
\end{equation}
with $[\tilde{\textbf{z}}_{i}] $ representing the \ac{AWGN} plus the error due to the phase-distortion removal procedure. 
In order to apply the \ac{OMP} method, a uniform grid of size $G$ is defined over the delay domain as $\mathfrak{G} = \{0, 1/G,..., G-1/G\}/f_{s}$, and a dictionary \textbf{D} as $\textbf{D} = [\textbf{d}(0),...,\textbf{d}(G-1)] \in \mathbb{C}^{MN\times G}$, where $G >\!\!> MN$ and each column $\textbf{d}(i)$ given as
\small
\begin{equation}
	\textbf{d}(i) = \dfrac{1}{\sqrt{MN}}[e^{-j2\pi[\textbf{f}]_{1}(\dfrac{i}{G})/f_{s}},...,e^{-j2\pi[\textbf{f}]_{MN}(\dfrac{i}{G})/f_{s}}]^{T} \in \mathbb{C}^{MN},
\end{equation}
\normalsize
where $i = 0,1,...,G-1$.
For values of $G = 2MN$, or $G = 3MN$ the grid $\mathfrak{G}$ is considered dense and the vector in~(\ref{eq:ss_received_pilots}) can be approximated to
\begin{equation}
	\tilde{\textbf{y}} \approx \textbf{D}\textbf{h}_{0} + \tilde{\textbf{z}}
\end{equation}
where $\textbf{h}_{0} \in \mathbb{C}^{G}$ is a discrete approximation for $h_{0}$, and is estimated using the \ac{OMP} sparse recovery method and the given $\tilde{\textbf{y}}$ samples.
The \ac{OMP} method is a greedy iterative algorithm, that selects a column of the dictionary \textbf{D}, at each iteration, such that it has the highest correlation with the current residual and it repeats until a convergence condition is met~\cite{tropp2007signal}. 
For each selected column, the non-zero coefficients are computed using the least-square method, such that they approximate the measurement vector $\tilde{\textbf{y}}$.
The algorithm stops once the number of the selected dictionary columns reaches the sparsity order of $\textbf{h}_{0}$, which is given as input.

%

\subsection{Practical Channel Splicing for OFDM Systems}
\label{sec:system}

In the following, we describe the system architecture and the estimation technique used for generating the \ac{CIR} and \ac{CFR}.
The \ac{SDR}-based channel sounder builds upon USRP-components to perform the over-the-air communication, and \MATLAB{} for the software implementation.

The \ac{OFDM} transmitter is implemented according to the 802.11ac standard. 
This standard support the signal bandwidths of \SIlist{20;40;80;160}{\mega\hertz}. 
Despite the different signal bandwidths, the \ac{OFDM} subcarrier spacing is kept fixed to \SI{312.5}{\kilo\hertz}, whereas the number of the subcarriers changes accordingly.
Firstly, the transmitter converts the payload message into a sequence of bits, which later are \ac{BPSK} and \ac{OFDM} modulated. 
For over-the-air experiments, the generated baseband signal is forwarded to the USRP \ac{SDR}, which upconverts it to the carrier frequency of interest. 
The connection between \MATLAB{} and the USRP is realized through the recently released Wireless Testbench toolbox\footnote{https://de.mathworks.com/products/wireless-testbench.html}, which supports high-speed data transmission.

We also use preamble detector functionality provided by the toolbox, allowing capturing only the signal of interest for offline analysis.
Once packets in the air are detected by the preamble detector, the signal is captured, downconverted, and the raw data is stored into a binary file for further post-processing on the host computer. 
During post-processing, the time and carrier frequency offset are estimated using the frame preamble and compensated per each received packet. 
The channel is estimated based on the \ac{VHT-LTF} sequence in the preamble using the \ac{LS} estimation technique, and the obtained \ac{CIR} and \ac{CFR} are stored.
Finally, the signal is decoded and demodulated, and the transmitted payload message is recovered.
The collected \ac{CFR} samples over multiple narrow frequency bands are used as input to the spectrum splicing technique for estimating the wideband channel.

Channel estimation is performed using \ac{LS} estimation time or frequency domain approach~\cite{heiskala2001ofdm} on the \ac{VHT-LTF}.
The time-domain approach computes the \ac{CIR} as
\begin{equation}\label{eq:cir_formula}
	\hat{h} = X^{\dagger} y
\end{equation}
where \textit{y} are the received samples, $\hat{h}$ is the \ac{CIR} and $X^{\dagger}$ is the Moore-Penrose (pseudo) inverse of the Toeplitz matrix $X$.
Likewise, the frequency-domain approach acquires the \ac{CFR} as 
\begin{equation}\label{eq:cfr_formula}
	\hat{H} = Y ./ X
\end{equation}
where \textit{Y} are the received samples, $\hat{H}$ is the \ac{CFR} and the \textit{X} are the transmitted samples. 
Both \ac{CIR} and \ac{CFR} can be computed from each other using the \ac{FFT} and \ac{IFFT}.

In the following, we describe the validation of our implementation in simulation as well as results from first over-the-air experiments in an indoor scenario.

%

\section{Performance Evaluation}
\label{sec:Evaluation}

Channel splicing aims to estimate a wideband channel by combining multiple narrow-band channel measurements. 
This work validates the splicing technique presented in~\cite{khalilsarai2020wifibased} in a controlled simulated environment and a real-world scenario, exploiting the flexibility of the developed tool to switch between theory and practice.
Currently, the validation focuses only on the accuracy of the estimated delay, ignoring the amplitude, which will be analyzed in future work.

\subsection{Simulations}

Splicing is validated in simulations at the frequency band \SI{5}{\giga\hertz} by splitting the wide  \SI{160}{\mega\hertz} bandwidth into narrower sub-bands, specifically $2 \times \SI{80}{\mega\hertz}$, $4 \times \SI{40}{\mega\hertz}$, and $8 \times \SI{20}{\mega\hertz}$ sub-bands. 
Simulations are run at the central frequencies listed in \cref{table:center_frequencies}.

\begin{table}[b]
    \vspace{-1em}
	\caption{Simulation measurement center frequencies}
	\label{table:center_frequencies}
	\centering
	\begin{tabular}{rl}
		\toprule
		Bandwidth & Center frequency \\
		\midrule
		\SI{80}{\mega\hertz} &  \SIlist[list-units=single,list-final-separator = {, }, list-pair-separator= {, }]
  {4.96;5.04}{ \si{\giga\hertz}}\\
		\SI{40}{\mega\hertz} & \SIlist[list-units=single,list-final-separator = {, }, list-pair-separator= {, }]
  {4.94;4.98;5.02;5.06}{ \si{\giga\hertz}}\\
		\SI{20}{\mega\hertz} & 
		\SIlist[list-units=single,list-final-separator = {, }, list-pair-separator= {, }]
  {4.93;4.95;4.97;4.99;5.01;5.03;5.05;5.07}{ \si{\giga\hertz}}\\
		\bottomrule
	\end{tabular}
\end{table}

The generated \acp{CFR} are combined into an array and used as input to the spectrum splicing technique. 
The estimated \ac{CIR} is compared towards the full \SI{160}{\mega\hertz} \ac{CIR} obtained utilizing the \ac{LS} method.
In the simulated environment, we up-/down-convert to the carrier frequency of \SI{5}{\giga\hertz} in software and make use of a the \MATLAB{} Rayleigh channel.
The channel model allows defining the number of the \acp{MPC} and the time delay for each path.
We investigate the algorithm performance in two scenarios, and with different number of multipath components and sub-bands width.

\subsubsection{Different Frequency Resolution}

\begin{figure}
\vspace{.5em}
	\centering
	\includegraphics[width=\linewidth]{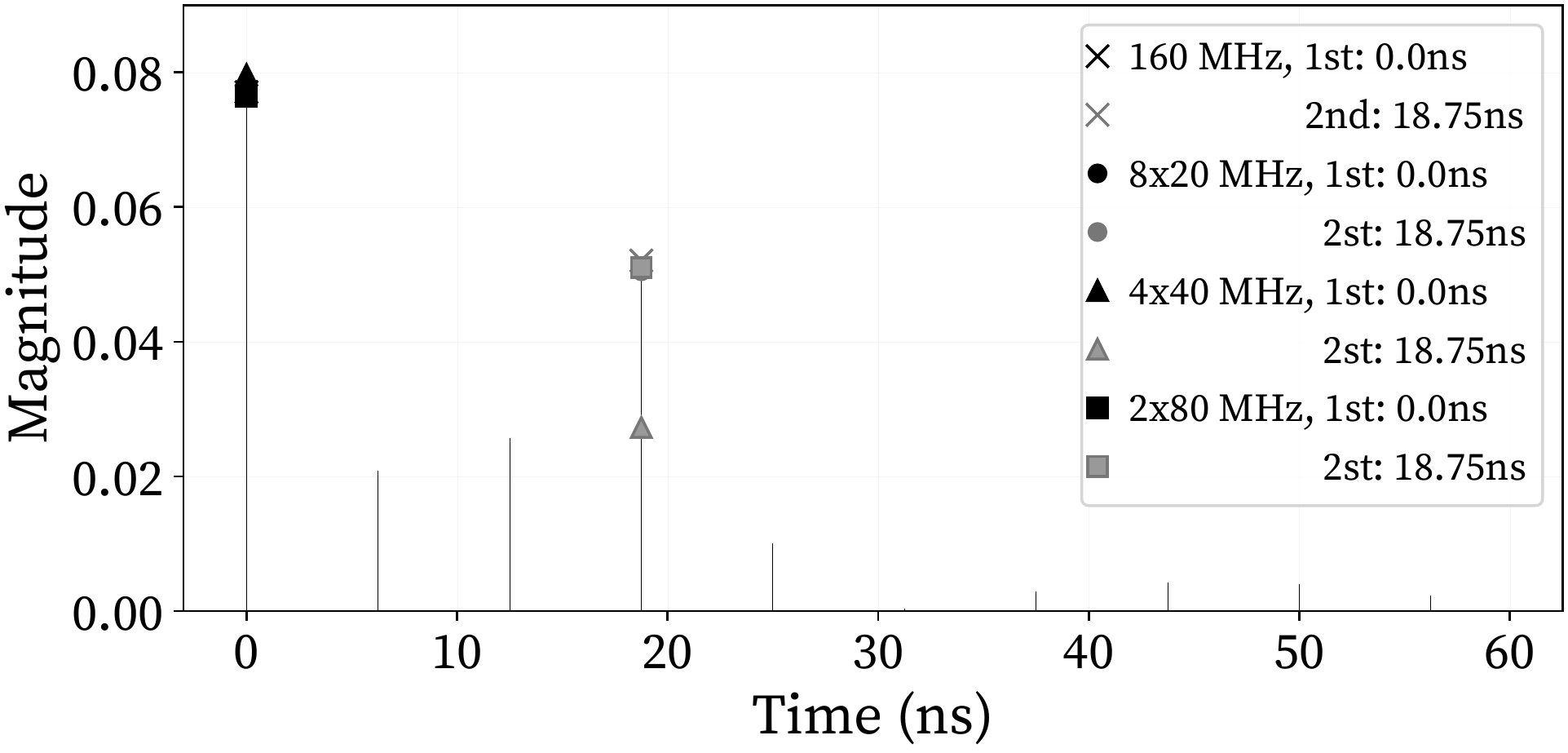}
	\caption{Estimated peaks using different sub-bands width compared towards the \SI{160}{\mega\hertz} \ac{CIR}.}
	\label{fig:allBands_fullData}
\end{figure}

\begin{figure}
	\centering
	\includegraphics[width=\linewidth]{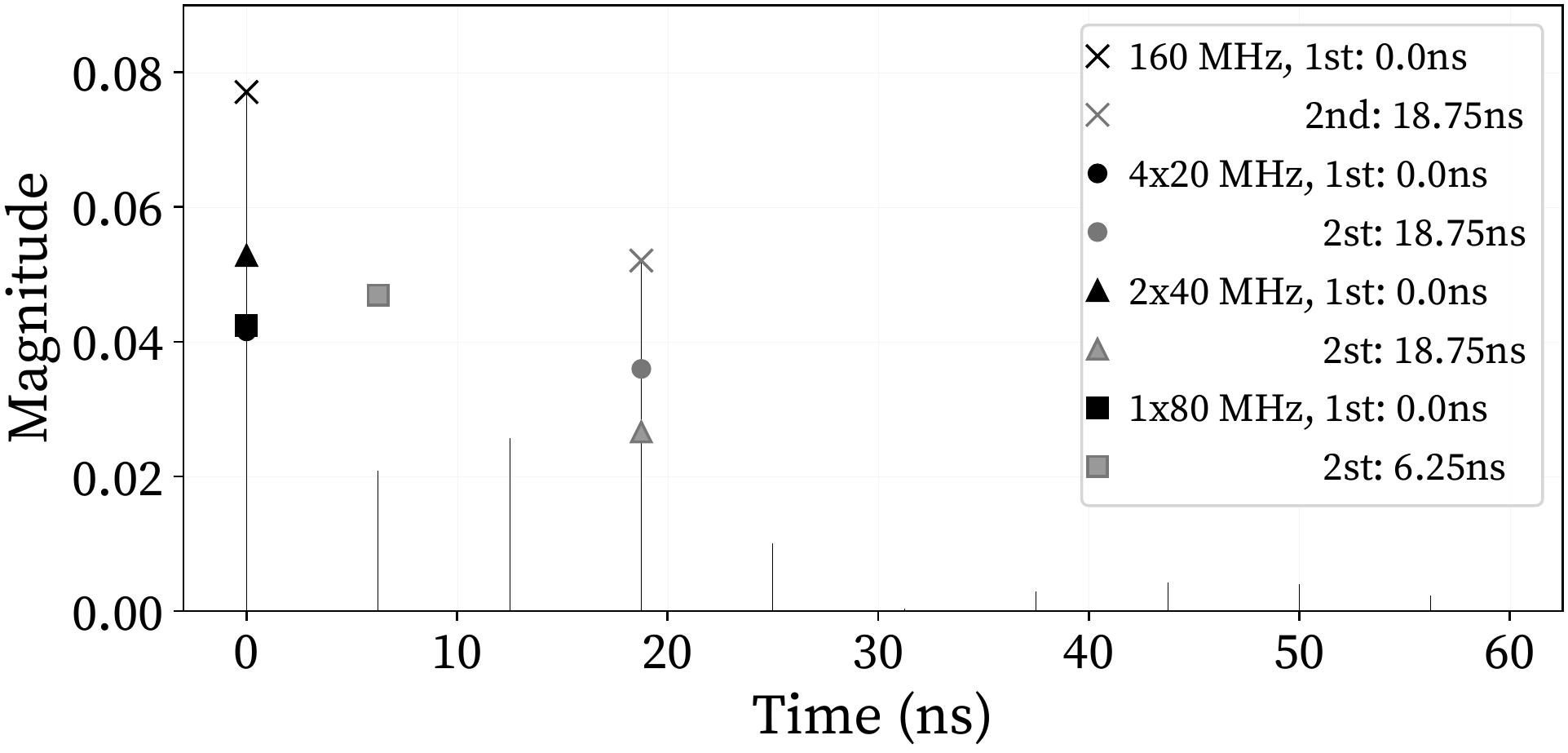}
	\caption{Estimated peaks using 50\% of the sub-bands, for different sub-bands width, compared towards \SI{160}{\mega\hertz} \ac{CIR}.}
	\label{fig:allBands_halData}
\vspace{-.5em}
\end{figure}

The first scenario consists of $\textit{K}=2$ \acp{MPC}, with delays and average powers set to $\{ 0, 18.75\}\si{\nano\second}$ and $\{ 0, -2\}\si{\decibel}$.
The splicing technique is applied over the collected \ac{CFR} samples for different sub-bands width at the corresponding center frequencies. 
The results are presented in \cref{fig:allBands_fullData} and depicts 
the estimated two paths (indicated by the markers) along with the computed \SI{160}{\mega\hertz} \ac{CIR}. 
Our technique correctly estimates the two \SI{160}{\mega\hertz} peaks despite the sub-channel width. 
In the considered scenario, the \ac{CFR} samples are collected over all the sub-bands. 
However, we further investigate the algorithm performance, looking at only half of the sub-bands. 
Similarly, the results are shown in \cref{fig:allBands_halData}.

As can be seen, the results for the $2 \times \SI{40}{\mega\hertz}$ and the $1 \times \SI{80}{\mega\hertz}$ show that we can correctly estimate the two strongest paths observed by the \SI{160}{\mega\hertz} \ac{CIR}, despite the missing samples. 
The results for the $4 \times \SI{20}{\mega\hertz}$ experiment, on the other hand, indicate that we fail estimating correctly the second path, with a difference up to 2 samples (1 sample = 1/160MHz = \SI{6.25}{\nano\second}).

\subsubsection{Multiple Paths}

\begin{figure}
\vspace{-.5em}
	\centering
	\includegraphics[width=\linewidth]{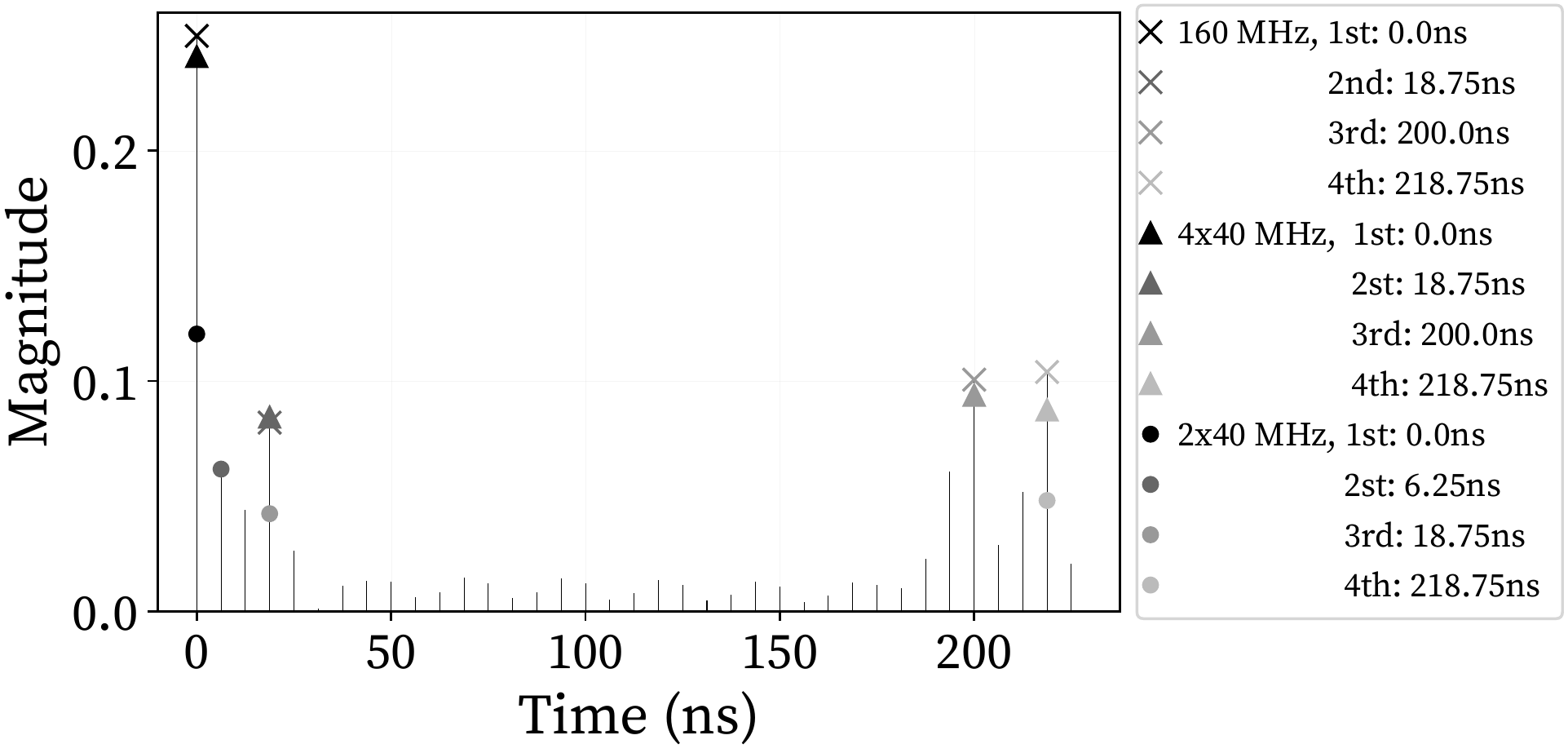}
	\caption{Estimated \acp{MPC} using spectrum splicing based on \SI{40}{\mega\hertz} sub-bands for the scenario with 4 paths.}
	\label{fig:scenario_4paths}
\end{figure}

In the second scenario, we consider $\textit{K}=4$ \acp{MPC}, with delays and average powers set to $\{ 0, 18.75, 200, 218.75\}\si{\nano\second}$ and $\{ 0, 0,  -2, 0\}\si{\decibel}$.
Splicing is applied on \acp{CFR} samples collected over all and $50\%$ of the \SI{40}{\mega\hertz} sub-bands, with the purpose to estimate the \SI{160}{\mega\hertz} \ac{CIR}. 
The results are presented in \cref{fig:scenario_4paths} and illustrate that splicing applied over all the sub-bands correctly estimate the $\textit{K}=4$ peaks delay, shown by the markers. 
Yet, the algorithm fails to estimate the third peak when only $50\%$ of the sub-bands are used.

To summarize, the simulation results show that, channel splicing correctly estimates the paths delay which are distinguishable only by a wide channel, in case it is applied over the \ac{CFR} samples collected over all the narrower sub-bands.
The algorithm performance in some cases deteriorate when $50\%$ of the sub-bands are utilized. 
As future work we aim to improve the algorithm performance when only $50\%$ of the sub-bands are utilized which would speed up the estimation of wider channels in the range of, e.g., \SI{1}{\giga\hertz}.

\subsection{Indoor Lab-Experiments}

\begin{table}[b]
    \vspace{-1em}
	\caption{Experimental measurement center frequencies}
	\label{table:eperimental_center_frequencies}
	\centering
	\begin{tabular}{rl}
		\toprule
		Bandwidth & Center frequency \\
		\midrule
		\SI{40}{\mega\hertz} &  \SIlist[list-units=single,list-final-separator = {, }, list-pair-separator= {, }]
  {2.43;2.47}{ \si{\giga\hertz}}\\
		\SI{20}{\mega\hertz} & \SIlist[list-units=single,list-final-separator = {, }, list-pair-separator= {, }]
  {2.42;2.44;2.46;2.48}{ \si{\giga\hertz}}\\
		\bottomrule
	\end{tabular}
\end{table}

\begin{figure}
\vspace{.5em}
	\centering
	\includegraphics[width=\linewidth]{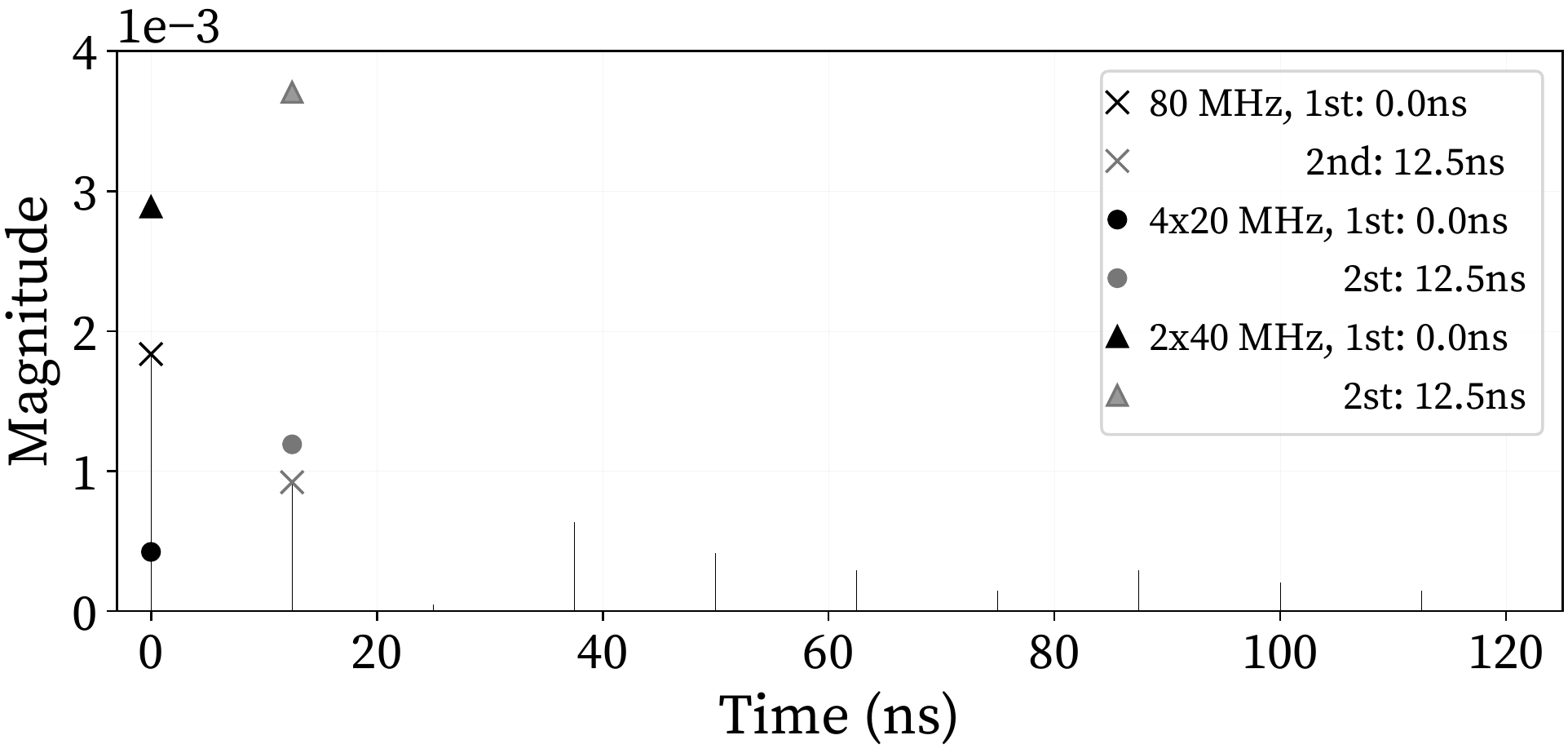}
	\caption{Estimated peaks from measurements conducted over all frequency bands and compared towards the \SI{80}{\mega\hertz} \ac{CIR}.}
	\label{fig:expRes_fullData}
\end{figure}

\begin{figure}
	\centering
	\includegraphics[width=\linewidth]{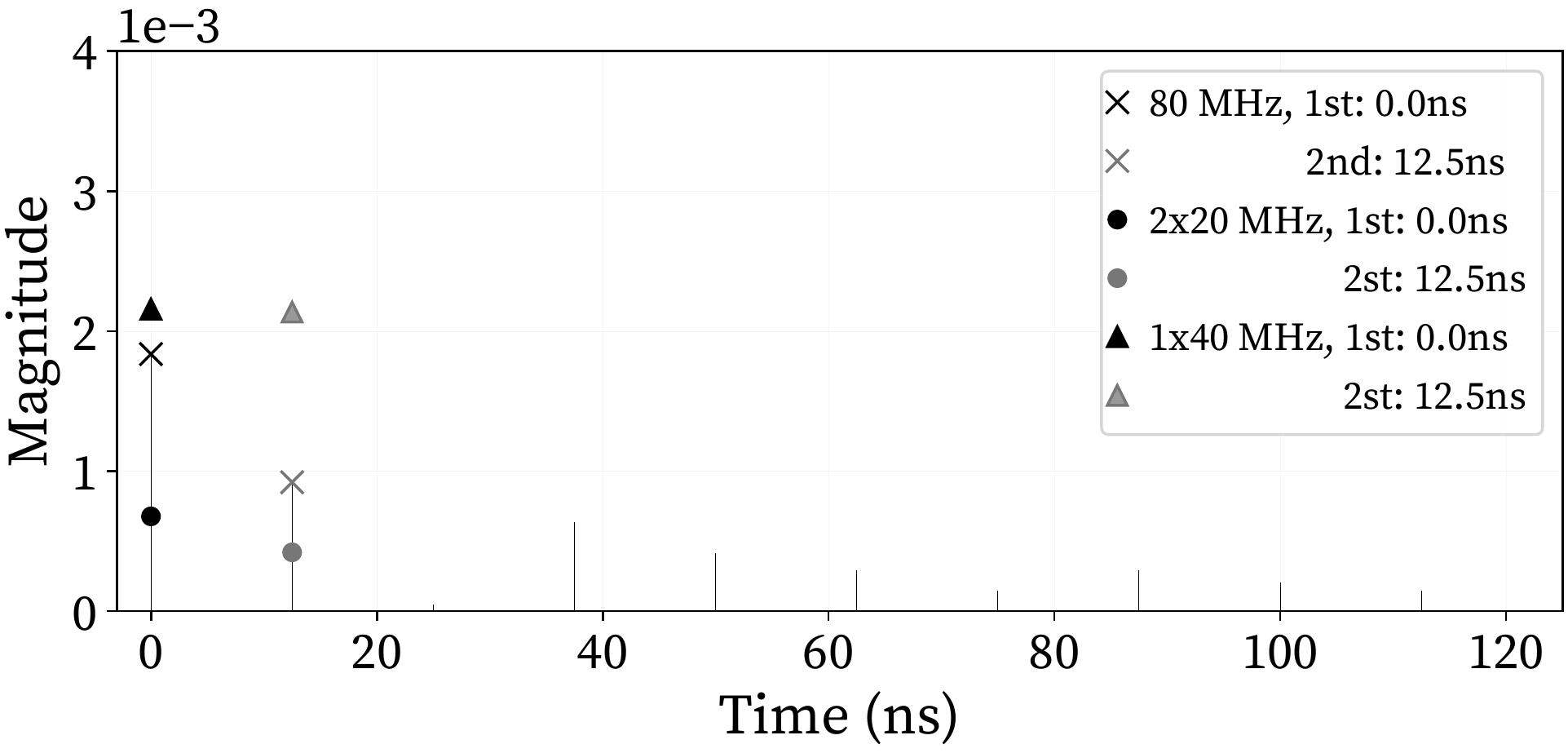}
	\caption{Estimated peaks from measurements conducted over 50\% of the sub-bands and compared towards the \SI{80}{\mega\hertz} \ac{CIR}.}
	\label{fig:expRes_halfData}
\vspace{-.5em}
\end{figure}

To verify the accuracy of the spectrum splicing technique, measurements are conducted in real-world scenarios.
Two USRP N310, correspondingly the transmitter and receiver, are connected through 1\,Gbps 
Ethernet to a host computer, in which two instances of Matlab are set up.
The experiments are performed indoor, at \SI{2.4}{\giga\hertz} using signal bandwidth of \SI{80}{\mega\hertz} and a communication distance of \SI{4}{\meter}.
Next, the wide bandwidth is divided into $2 \times \SI{40}{\mega\hertz}$, $4 \times \SI{20}{\mega\hertz}$ sub-bands and raw data is collected at the center frequencies listed in \cref{table:eperimental_center_frequencies}.

To estimate the wide bandwidth \ac{CIR}, spectrum splicing is applied on the collected \ac{CFR} samples over all and half of the sub-bands.
The results are presented in \cref{fig:expRes_fullData,fig:expRes_halfData}, respectively.
The plots depict the estimated two strongest paths for different sub-channel width (illustrated by the markers), compared towards the \SI{80}{\mega\hertz} \ac{CIR}. 
In all cases the method correctly compute the strongest \SI{80}{\mega\hertz} paths in terms of delay.

\begin{figure}
	\centering
	\includegraphics[width=\linewidth]{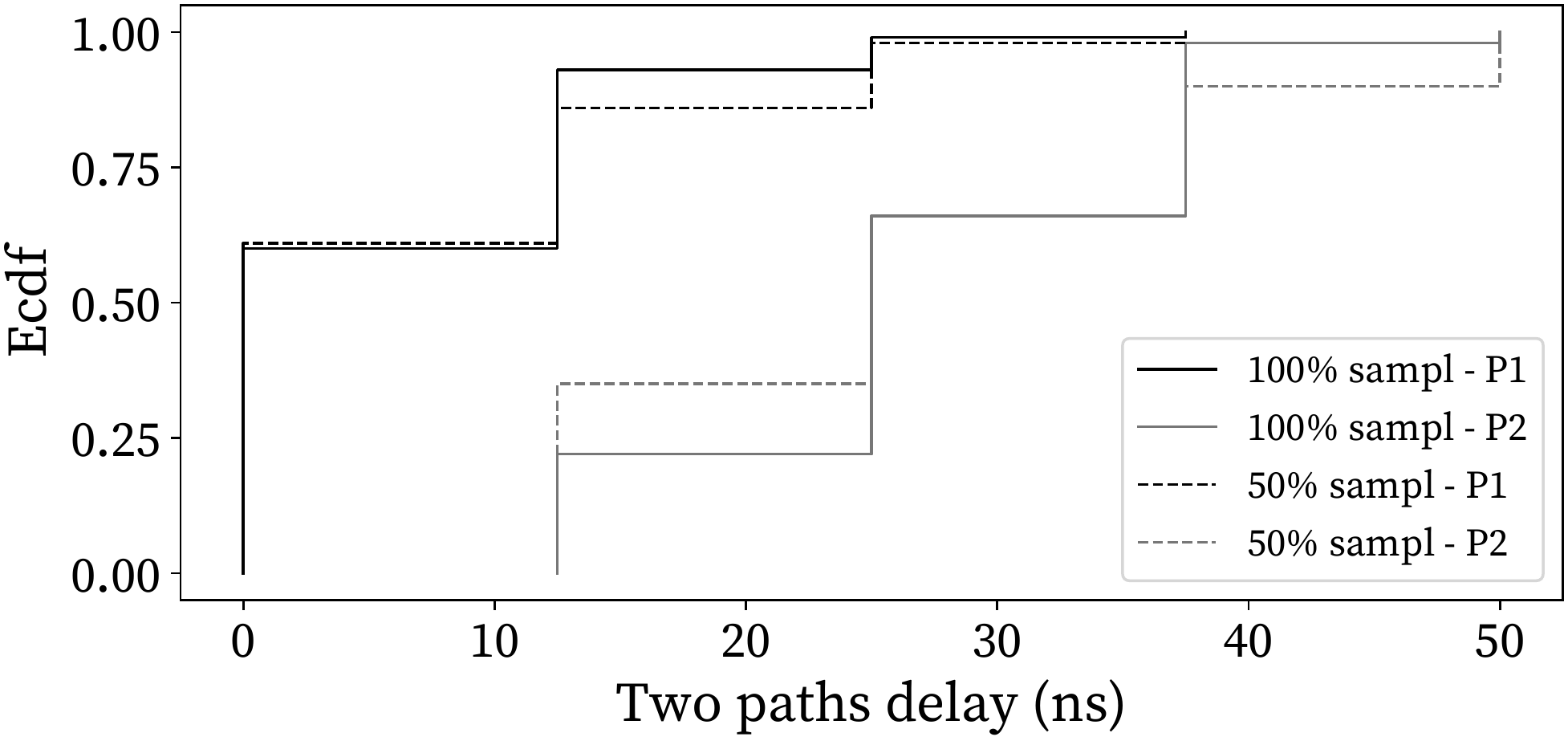}
	\caption{Computed ecdf for 2 paths estimated from splicing applied over \SI{20}{\mega\hertz} sub-bands over 100 packets.}
	\label{fig:ecdf}
\vspace{-.5em}
\end{figure}

However, differently from simulations, the amplitude of the \ac{CIR} slightly fluctuates over the collected packets at each band as a consequence of the dynamic environment, even though the transmitter and the receiver are static. 
This slight fluctuation impacts the \ac{OMP} algorithm on delay estimation.

Therefore, to study the algorithm performance, we collected 100 packets and computed the ecdf for the estimated two strongest paths for the 100 packets, over all and 50\% \SI{20}{\mega\hertz} sub-bands.
The ground truth are the two \SI{80}{\mega\hertz} peaks delay, respectively, \SI{0}{\nano\second} and \SI{12.5}{\nano\second}.
The results are given in \cref{fig:ecdf} and show that for such narrow bands, the method sometimes overestimate the delays. 
Considering that the resolution of the wide band is $1/\SI{80}{\mega\hertz} =\SI{12.5}{\nano\second}$, the first path is sometimes overestimated up to 2 samples, whereas the second path up two 3 samples. 
As a future work, we aim to optimize the method, such that the impact of the amplitude variation is ignored.


%

\section{Conclusion}
\label{sec:conclusion}

Following the key concepts of \acl{JCAS}, we realized an OFDM-based channel sounder using spectrum splicing.
Spectrum splicing allows to use measurements of multiple narrow-band subchannels to obtain precise wideband measurements.
In particular, we validated the low-complexity grid-based spectrum splicing algorithm in simulations and real-world-scenario -- making it feasible for \acf{IoT} solutions.
The algorithm is integrated into an IEEE 802.11ac based communication system, where \ac{CIR} and \ac{CFR} are estimated using the \ac{LS} estimation technique.
For indoor lab experiments, we implemented the system using an USRP-based \ac{SDR}.
We were particularly interested in the ability to obtain wideband channel properties using only a subset of the narrow-band subchannels.
Our system allows a very good estimation of the multipath  components in terms of time delay both in simulation as well as in indoor experiments.

We see this work as a first step towards measurements of even wider band channels in the mmWave bands.
In future work, we plan further experiments also covering longer distances and more dynamic scenarios.
In addition, we aim to analyze the algorithm performance in computing the amplitude for the estimated multi-path components.

\printbibliography

@article{cui2021integrating,
	author = {Cui, Yuanhao and Liu, Fan and Jing, Xiaojun and Mu, Junsheng},
	title = {{Integrating Sensing and Communications for Ubiquitous IoT: Applications, Trends, and Challenges}},
	doi = {10.1109/mnet.010.2100152},
	issn = {1558-156X},
	journal = {IEEE Network},
	month = Sep,
	number = {5},
	pages = {158--167},
	publisher = {IEEE},
	volume = {35},
	year = {2021},
}

@book{heiskala2001ofdm,
	author = {Heiskala, Juha and Terry, John},
	title = {{OFDM Wireless LANs: A Theoretical and Practical Guide}},
	isbn = {978-0-672-32157-3},
	location = {Indianapolis, IN},
	publisher = {SAMS},
	year = {2001},
}

@article{kazaz2021delay,
	author = {Kazaz, Tarik and Janssen, Gerard JM and Romme, Jac and Van der Veen, Alle-Jan},
	title = {{Delay estimation for ranging and localization using multiband channel state information}},
	doi = {10.1109/TWC.2021.3113771},
	issn = {1536-1276},
	journal = {IEEE Transactions on Wireless Communications (TWC)},
	number = {4},
	pages = {2591--2607},
	publisher = {IEEE},
	volume = {21},
	year = {2021},
}

@techreport{khalilsarai2020wifibased,
	author = {Khalilsarai, Mahdi Barzegar and Gross, Benedikt and Stefanatos, Stelios and Wunder, Gerhard and Caire, Giuseppe},
	title = {{WiFi-Based Channel Impulse Response Estimation and Localization via Multi-Band Splicing}},
	doi = {10.48550/arXiv.2011.10402},
	institution = {arXiv},
	month = Nov,
	number = {2011.10402},
	type = {cs.IT},
	year = {2020},
}

@inproceedings{le2021deep,
	author = {Le Ha, An and Van Chien, Trinh and Nguyen, Tien Hoa and Choi, Wan and Duc Nguyen, Van},
	title = {{Deep Learning-Aided 5G Channel Estimation}},
	booktitle = {15th International Conference on Ubiquitous Information Management and Communication (IMCOM 2021)},
	address = {Seoul, South Korea},
	doi = {10.1109/IMCOM51814.2021.9377351},
	month = Jan,
	pages = {1--7},
	publisher = {IEEE},
	year = {2021},
}

@inproceedings{liyanaarachchi2020joint,
	author = {Liyanaarachchi, Sahan Damith and Barneto, Carlos Baquero and Riihonen, Taneli and Valkama, Mikko},
	title = {{Joint OFDM Waveform Design for Communications and Sensing Convergence}},
	booktitle = {IEEE International Conference on Communications (ICC 2020)},
	address = {Virtual Conference},
	doi = {10.1109/icc40277.2020.9149408},
	issn = {1550-3607},
	month = Jun,
	publisher = {IEEE},
	year = {2020},
}

@article{lv2019channel,
	author = {Lv, Changwei and Lin, Jia-Chin and Yang, Zhaocheng},
	title = {{Channel prediction for millimeter wave MIMO-OFDM communications in rapidly time-varying frequency-selective fading channels}},
	issn = {2169-3536},
	journal = {IEEE Access},
	month = Jan,
	pages = {15183--15195},
	publisher = {IEEE},
	year = {2019},
}

@book{rappaport1996wireless,
	author = {Rappaport, Theodore S.},
	title = {{Wireless Communications: Principles and Practice}},
	location = {Upper Saddle River, NJ},
	publisher = {Prentice Hall},
	year = {1996},
}

@article{shen2019wirim,
	author = {Shen, Xinbin and Guo, Lingchao and Lu, Zhaoming and Wen, Xiangming and He, Zhihong},
	title = {{WiRIM: Resolution improving mechanism for human sensing with commodity Wi-Fi}},
	issn = {2169-3536},
	journal = {IEEE Access},
	month = Nov,
	pages = {168357--168370},
	publisher = {IEEE},
	volume = {7},
	year = {2019},
}

@inproceedings{sun2017design,
	author = {Sun, Ruoyu and Papazian, Peter B and Senic, Jelena and Lo, Yeh and Choi, Jae-Kark and Remley, Kate A and Gentile, Camillo},
	title = {{Design and calibration of a double-directional 60 GHz channel sounder for multipath component tracking}},
	booktitle = {11th European Conference on Antennas and Propagation (EUCAP 2017)},
	address = {Paris, France},
	month = Mar,
	pages = {3336--3340},
	publisher = {IEEE},
	year = {2017},
}

@article{tropp2007signal,
	author = {Tropp, Joel A and Gilbert, Anna C},
	title = {{Signal recovery from random measurements via orthogonal matching pursuit}},
	issn = {0018-9448},
	journal = {IEEE Transactions on Information Theory},
	number = {12},
	pages = {4655--4666},
	publisher = {IEEE},
	volume = {53},
	year = {2007},
}

@inproceedings{vasisht2016decimeter,
	author = {Vasisht, Deepak and Kumar, Swarun and Katabi, Dina},
	title = {{Decimeter-level localization with a single WiFi access point}},
	booktitle = {13th USENIX Symposium on Networked Systems Design and Implementation (NSDI 2016)},
	address = {Santa Clara, CA},
	month = Mar,
	pages = {165--178},
	year = {2016},
}

@techreport{wan2022fundamental,
	author = {Wan, Yubo and Liu, An and Du, Rui and Han, Tony Xiao},
	title = {{Fundamental Limits and Optimization of Multiband Sensing}},
	doi = {10.48550/arXiv.2207.10306},
	institution = {arXiv},
	month = Jul,
	type = {eess.SP},
	year = {2022},
}

@techreport{wan2022multiband,
	author = {Wan, Yubo and Liu, An and Hu, Qiyu and Zhang, Mianyi and Cai, Yunlong},
	title = {{Multiband Delay Estimation for Localization Using a Two-Stage Global Estimation Scheme}},
	institution = {arXiv},
	month = Jun,
	type = {eess.SP},
	year = {2022},
}

@article{wang2022integrated,
	author = {Wang, Jian and Varshney, Neeraj and Gentile, Camillo and Blandino, Steve and Chuang, Jack and Golmie, Nada},
	title = {{Integrated Sensing and Communication: Enabling Techniques, Applications, Tools and Datasets, Standardization, and Future Directions}},
	issn = {2327-4662},
	journal = {IEEE Internet of Things Journal},
	month = Jul,
	publisher = {IEEE},
	year = {2022},
}

@article{wild2021joint,
	author = {Wild, Thorsten and Braun, Volker and Viswanathan, Harish},
	title = {{Joint Design of Communication and Sensing for Beyond 5G and 6G Systems}},
	doi = {10.1109/ACCESS.2021.3059488},
	issn = {2169-3536},
	journal = {IEEE Access},
	month = Jan,
	publisher = {IEEE},
	volume = {9},
	year = {2021},
}

@article{wu2021otfsbased,
	author = {Wu, Kai and Zhang, J. Andrew and Huang, Xiaojing and Guo, Y. Jay},
	title = {{OTFS-Based Joint Communication and Sensing for Future Industrial IoT}},
	doi = {10.1109/jiot.2021.3139683},
	issn = {2327-4662},
	journal = {IEEE Internet of Things Journal},
	month = Feb,
	number = {3},
	publisher = {IEEE},
	volume = {10},
	year = {2023},
}

@article{xie2018precise,
	author = {Xie, Yaxiong and Li, Zhenjiang and Li, Mo},
	title = {{Precise Power Delay Profiling with Commodity Wi-Fi}},
	doi = {10.1109/TMC.2018.2860991},
	issn = {1536-1233},
	journal = {IEEE Transactions on Mobile Computing (TMC)},
	month = Jun,
	number = {6},
	pages = {1342--1355},
	publisher = {IEEE},
	volume = {18},
	year = {2019},
}

@article{zhang2022enabling,
	author = {Zhang, J. Andrew and Rahman, Md. Lushanur and Wu, Kai and Huang, Xiaojing and Guo, Y. Jay and Chen, Shanzhi and Yuan, Jinhong},
	title = {{Enabling Joint Communication and Radar Sensing in Mobile Networks - A Survey}},
	doi = {10.1109/comst.2021.3122519},
	issn = {1553-877X},
	journal = {IEEE Communications Surveys \& Tutorials},
	number = {1},
	pages = {306--345},
	publisher = {IEEE},
	volume = {24},
	year = {2022},
}

\end{document}